\documentclass[11pt]{article}
\newcommand{\ket}[1]{\left | #1 \right \rangle}
\newcommand{\bra}[1]{\left \langle #1 \right |}

\def\openone{\leavevmode\hbox{\small1\kern-3.8pt\normalsize1}}
\def\RR{{\rm I\kern-.2emR}}
\def\tr{{\rm tr}\; }
\def\ce{{\cal E}}

\def\cb{{\cal B}}
\def\ch{{\cal H}}

\def\ci{{\cal I}}
\def\ca{{\cal A}}

\def\cp{{\cal P}}

\newtheorem{theorem}{Theorem}
\newtheorem{definition}{Definition}
\newtheorem{lemma}{Lemma}

\newtheorem{proposition}{Proposition}

\newcommand{\proj}[1]{\ket{#1}\!\bra{#1}}

\newcommand{\inner}[2]{ \langle #1 | #2 \rangle}

\newcommand{\beq}{\begin{equation}}
\newcommand{\eeq}{\end{equation}}
\newcommand{\beqa}{\begin{eqnarray}}
\newcommand{\eeqa}{\end{eqnarray}}

\newcommand{\qed}{\mbox{\rule[0pt]{1.5ex}{1.5ex}}}

\begin{document}
\begin{center}
{\Large\bf On the reversible extraction of classical information
\\from a quantum source }\\
\bigskip
{\normalsize Howard Barnum$^\dagger$, Patrick Hayden$^*$, Richard
Jozsa$^\dagger$, and Andreas Winter$^\S$}\\
\bigskip
{\small\it $^\dagger$Department of Computer Science, University of
Bristol,\\ Merchant Venturers Building, Bristol BS8 1UB U.K. \\

$^*$Centre for Quantum Computation, Clarendon Laboratory,\\ Parks
Road, Oxford OX1 3PU, U.K.\\

$^\S$SFB 343, Facult\"{a}t f\"{u}r Mathematik, Universit\"{a}t
Bielefeld, \\Postfach 100131, 33501 Bielefeld, Germany.}
\\[4mm]
\date{today}
\end{center}

\begin{abstract}
Consider a source $\ce$ of pure quantum states with von Neumann
entropy $S$. By the quantum source coding theorem, arbitrarily long
strings of signals may be encoded asymptotically into $S$
qubits/signal (the Schumacher limit) in such a way that entire
strings may be recovered with arbitrarily high fidelity. Suppose that
classical storage is free while quantum storage is expensive and
suppose that the states of $\ce$ do not fall into two or more
orthogonal subspaces. We show that if $\ce$ can be compressed with
arbitrarily high fidelity into $A$ qubits/signal plus any amount of
auxiliary classical storage then $A$ must still be at least as large
as the Schumacher limit $S$ of $\ce$. Thus no part of the quantum
information content of $\ce$ can be faithfully replaced by classical
information. If the states do fall into orthogonal subspaces then $A$
may be less than $S$, but only by an amount not exceeding the amount
of classical information specifying the subspace for a signal from
the source.

\end{abstract}

\section{Introduction}\label{sect1}
The quantum source coding theorem
\cite{Schumacher95a,Jozsa94a,Barnum96b,Winter99,JHHH} provides one of
the clearest manifestations of the concept of quantum information. It
characterises the minimal resource (in terms of Hilbert space
dimension) that is sufficient to faithfully represent long sequences
of signal emissions from a memoryless quantum source. This provides a
notion of the quantum information content of the source and the
minimal resource is given by the Schumacher limit -- $S$
qubits/signal -- where $S$ is the von Neumann entropy of the source.
In this paper we consider a possible refinement of this theorem,
asking to what extent the quantum information may be represented in
two parts -- a classical part and a quantum part -- such that the
quantum part is minimised while the classical part may be as large as
desired. We will show that it is impossible to reduce the resource of
the quantum part to below the Schumacher limit, except in the special
case that the signal states fall into two or more orthogonal
subspaces. Thus in general (i.e. with the preceding exception) it is
impossible to substitute classical information for any part of the
quantum information of a source.

The paper is organised as follows. The main results are given in
theorems \ref{3} and \ref{5} of section \ref{sect4}. We approach
the proofs of these results through a sequence of lemmas after
establishing some preliminary definitions and terminology. In
section \ref{sect2} we provide a formal definition of a
coding--decoding scheme which applies to blocks of signals of
general length $n$. We define the fidelity of any such scheme and
state Schumacher's quantum source coding theorem. In section
\ref{sect3} we introduce the distinction between reducible and
irreducible sources i.e. sources whose signal states respectively
do or do not fall into orthogonal subspaces. This distinction is
fundamental for our main results and we give an alternative
characterisation of it which is used in our subsequent proofs.

In section \ref{sect4} we refine the concept of coding--decoding
schemes to a situation in which the encoding has a classical part
and a quantum part. In terms of this concept we briefly review
earlier work of \cite{oldrevext} which provided the motivation of
our present study, and we give a precise statement of our main new
results. For any such refined coding--decoding scheme the
classical part of the encoding may be assumed to remain intact
after the input signal blocks have been reconstructed (with some
fidelity) by decoding. Correspondingly in section \ref{sect5} we
begin the proof of our main results by considering the classical
mutual information $\ci$ between the identity of the input string
and the classical part of the encoding. If the input string has
length $n$ and $Q$ denotes the number of qubits needed to support
the quantum part of the encoding, then we prove that $(Q+ \ci )/n$
cannot remain less than the Schumacher limit as the fidelity of
the coding--decoding scheme approaches unity. Finally to provide a
lower bound for $Q/n$, in section \ref{sect6} we study the
behaviour of $\ci/n$ for irreducible and reducible sources. We
prove that $\ci/n$ must tend to zero for any irreducible source as
the fidelity of the coding--decoding scheme tends to unity. For
reducible sources the situation is more complicated: clearly it is
possible to at least determine the identity of the orthogonal
subspace to which a given signal belongs, without disturbing the
signal. We prove that this is the best we can do i.e. that $\ci/n$
cannot exceed the amount of classical information about the signal
provided by this identification.

These lemmas in section \ref{sect6} are mathematically precise
examples of a heuristic principle in quantum information theory
viz. that it is impossible to obtain information about the
identity of a quantum state from an irreducible source without
irreparably disturbing the state and furthermore, that there
should be a trade-off between the amount of disturbance and the
amount of information gained. Such information--disturbance
results have been derived in other situations \cite{fuchs} but for
us there are extra technical complications arising from the fact
that the block length $n$ must generally increase unboundedly as
the fidelity of the coding--decoding scheme tends to unity i.e. we
have a situation in which the source varies as the disturbance
tends to zero. Our lemmas in section \ref{sect6}, referring to a
situation of unboundedly increasing block lengths, may have a
wider applicability for example to the study of the security of
quantum cryptographic protocols, in which an eavesdropper may
attempt to extract classical information from blocks of signal
transmissions.

In section \ref{sect7} we draw together the lemmas of the
preceding sections to give proofs of our main results. Finally in
section \ref{sect8} we summarise our findings and discuss some
related open questions.

\section{Preliminary definitions} \label{sect2}

We begin with a more precise statement of the quantum source
coding theorem which will also serve to establish terminology and
notations for our main results. We sometimes denote the ensemble
or source of (generally mixed) states $\xi_i$ with prior
probabilities $p_i$ as $\{ \xi_i ; p_i \}$. Consider a source $\ce
= \{ \ket{\sigma_i} ; p_i \} $ of pure quantum signal states
$\ket{\sigma_i}$ with prior probabilities $p_i$. We will use
capital letter indices to denote multi-indices for blocks of
signals of length $n$: \beqa \ket{\sigma_I}& =& \ket{\sigma_{i_1}}
\otimes \ldots \otimes \ket{\sigma_{i_n}} \\ p_I & = & p_{i_1}
\ldots p_{i_n} \\ I & = & i_1 \ldots i_n \eeqa We will often write
the projector $\proj{\sigma_I}$ simply as $\sigma_I$. Let $\ch$
denote the Hilbert space of single signals, of dimension $k$, and
let $\cb_\alpha$ denote the space of all mixed states of $\alpha$
qubits (or the smallest integer greater than $\alpha$ if $\alpha$
is not an integer). Then $n$-strings $\sigma_I$ are in
$\ch^{\otimes n}$ and in $\cb_{n\log k}$. In this paper,
logarithms are always to base 2.

If $\ket{\psi}$ and $\rho$ are any pure and mixed state
respectively in the same state space, we define the fidelity $F$
by \beq F(\proj{\psi}, \rho ) = \bra{\psi} \rho \ket{\psi} \eeq
More generally if $\omega$ and $\rho$ are mixed states we define
the fidelity by \cite{Uhlmann76a,Jozsa94b} \beq F(\rho ,\omega ) =
\left( \tr \sqrt{\sqrt{\omega}\rho \sqrt{\omega}} \right)^2 \eeq
The von Neumann entropy $S$ of $\ce$ is defined by \beq S= -\tr
\rho \log \rho \eeq where $\rho = \sum_i p_i \proj{\sigma_i}$ is
the overall density matrix of the signal states.

An encoding-decoding scheme for blocks of length $n$, to $\alpha$
qubits/signal and average fidelity $1-\epsilon$, is defined by the
following ingredients:\\ (i) an encoding operation $E_n :
\ch^{\otimes n} \rightarrow \cb_{n\alpha}$ which is a completely
positive trace preserving map (a CPTP map).\footnote{These encoding
operations are called blind, in contrast to visible encodings in
which $E_n$ is allowed to be an arbitrary map, but $D_n$ in (ii) is
still required to be CPTP. See \cite{mixed} for a further discussion
of this distinction. Note that the visible situation is trivial for
our main problem: all of the information about the input may be
faithfully extracted in classical form by recording the identity of
the input labels.} $E_n (\sigma_I) $ is a (mixed) state of $n\alpha$
qubits called the encoded or compressed version of $\sigma_I$.\\ (ii)
a decoding operation $D_n : \cb_{n\alpha} \rightarrow \cb_{n\log k} $
which is also a CPTP map. We write $\tilde{\sigma}_I = D_n E_n
(\sigma_I ) $ and call it the decoded version of $\sigma_I$. Note
that $\tilde{\sigma}_I$ is generally a mixed state.\\ (iii) the
average fidelity between the $\sigma_I$'s and $\tilde{\sigma}_I$'s is
$1-\epsilon$: \beq \sum_I p_I F(\sigma_I , \tilde{\sigma}_I ) =
1-\epsilon \eeq

We say that the source $\ce$ may be compressed to $\alpha$
qubits/signal if the following condition is satisfied: for all
$\epsilon > 0$ there is an $n_0$ such that for all block lengths
$n>n_0$ there is an encoding-decoding scheme for blocks of length $n$
to $\alpha$ qubits/signal and average fidelity at least $1-\epsilon$.
We can now make the following precise statement.

{\bf Quantum source coding theorem.}
\cite{Schumacher95a,Jozsa94a,Barnum96b,Winter99} Let $S$ be the
von Neumann entropy of a source $\ce$ of pure quantum states and
suppose that $\alpha \neq S$. Then $\ce$ may be compressed to
$\alpha$ qubits/signal if and only if $\alpha
>S$. $\qed$

\section{Reducible and irreducible sources} \label{sect3}

For our main results it will be important to classify sources
according to whether or not they decompose into orthogonal parts
in the following sense:

\begin{definition}  A source $\ce$ is called reducible if its signal states fall
into two or more orthogonal subspaces. Otherwise $\ce$ is called
irreducible. \end{definition}

If $\ce_i =\{ \alpha_{ij};p_{ij} \}_j$ for $i=1, \ldots ,L$ are
sources of signals lying in mutually orthogonal subspaces then we
may construct the reducible source $\ce = \bigcup_i a_i \ce_i = \{
\alpha_{ij} ; a_i p_{ij} \}_{ij}$ where $\{ a_1 , \ldots ,a_L \}$
is any chosen probability distribution (and the subscript outside
the bracket is the index labelling the signal states). Conversely
any reducible source $\ce$ may be decomposed into irreducible
parts $\ce = \bigcup a_i \ce_i$ by choosing a maximal orthogonal
decomposition. Here $a_i$ is the total probability of all states
of $\ce$ lying in the $i^{\rm th}$ orthogonal subspace and $\ce_i$
comprises these states with suitably renormalised probabilities.

We give an alternative characterisation of irreducibility of a
source of pure states which will be used in our later proofs.

\begin{definition} If $\ket{\sigma}$ and $\ket{\tau}$ are any signal states, a
chain from $\ket{\sigma}$ to $\ket{\tau}$ of length $m$ is a
sequence of signal states $\ket{\sigma_i}$ beginning with
$\ket{\sigma}$ and ending with $\ket{\tau}$:
\[ \ket{\sigma}=\ket{\sigma_1}, \ket{\sigma_2} \ldots
,\ket{\sigma_m}=\ket{\tau} \] such that
$\inner{\sigma_i}{\sigma_{i+1}} \neq 0$ for all $i=1, \ldots
,m-1$.\end{definition}

\begin{lemma} \label{7} Let $\ce$ be an ensemble with $K$ signal states.\\ (a) $\ce$
is irreducible {\em if and only if} for any two signal states
$\ket{\sigma}$ and $\ket{\tau}$ there is a chain from
$\ket{\sigma}$ to $\ket{\tau}$.\\ (b) If there is a chain from
$\ket{\sigma}$ to $\ket{\tau}$ then there is a chain of length at
most $K$.\end{lemma}

{\bf Proof} We will prove the contrapositive form of (a). Thus
suppose that the signal states do fall into two orthogonal
subspaces $E_1$ and $E_2$. Let $\ket{\sigma} \in E_1$ and
$\ket{\tau} \in E_2$. Then any chain from $\ket{\sigma}$ to
$\ket{\tau}$ would have a jump from $E_1$ to $E_2$ at some stage.
But this is impossible so there can be no chain from
$\ket{\sigma}$ to $\ket{\tau}$. Conversely suppose that there is
no chain from $\ket{\sigma}$ to $\ket{\tau}$. Let $M_{\sigma}$ be
the set of all signal states that are reachable from
$\ket{\sigma}$ by chains. Let $M^{c}_{\sigma}$ be the complement.
Thus $\ket{\sigma}\in M_\sigma$ and $\ket{\tau} \in M^c_\sigma$,
so both sets are non-empty. Now any $\ket{\tau'}\in M^c_\sigma$ is
orthogonal to all signals in $M_\sigma$ (since if
$\inner{\sigma'}{\tau'}\neq 0$ for some $\ket{\sigma'}\in
M_\sigma$ we would have a chain from $\ket{\sigma}$ to
$\ket{\sigma'}$ that extends to $\ket{\tau'}$, which is
impossible). Let $E_1$ and $E_2$ be the linear span of signals in
$M_\sigma$ and $M^c_\sigma$ respectively. Then $E_1$ and $E_2$ are
orthogonal subspaces containing all the signal states i.e. $\ce$
is reducible.\\ (b) Suppose that a chain from $\ket{\sigma}$ to
$\ket{\tau}$ contains some signal $\ket{\sigma'}$ twice:
\[ \ket{\sigma}, \ldots , \ket{\sigma'}, \ldots , \ket{\sigma'},
\ldots \ket{\tau}. \] Then we may delete the section between the
two $\ket{\sigma'}$'s and still have a chain. Hence if there is a
chain there is also a chain that contains each signal at most once
i.e. having length at most $K$.\, $\qed$ \\ {\bf Example} $\ce$
may contain orthogonal states yet still be irreducible. Minimal
chains may need to have maximal length $K$. Consider for example
$\ce$ with $K=5$ states given by $\ket{0}, \ket{0}+\ket{1},
\ket{1}+\ket{2}, \ket{2}+\ket{3}, \ket{3}$. Then $\ce$ is
irreducible. $\ket{0}$ is orthogonal to $\ket{3}$ and the shortest
chain between them has 5 members. \, $\qed$

\section{Coding with a classical and quantum part} \label{sect4}

In the context of quantum information theory, classical
information may be thought of as a special case {\em viz} the
quantum information of a source of states that are known (or
required) to always be members of a prescribed orthonormal basis.
More generally we may consider a quantum register as holding only
classical information (relative to a prescribed orthonormal basis)
if there is an omni-present fully decohering operation acting on
the register, diagonal in the basis, which prevents the occurrence
of any non-trivial superpositions of the basis states or any
entanglements of this register with any other quantum registers
being considered. Thus the most general allowable (``classical'')
state of the register is a probabilistic mixture of the basis
states, which may be classically correlated to the quantum  state
of all other registers (cf eq. (\ref{state}) below). These
conditions endow classical information with special properties not
shared by quantum information in general. For example classical
information is robust compared to quantum information -- it may be
readily stabilised and corrected by frequent measurement in the
given basis, which would destroy genuine quantum information.
Also, unlike quantum information, it may be cloned or copied.
These and other singular properties indicate that for many
purposes it is useful to regard classical information as a
separate resource, distinct from quantum information. In this
vein, it is natural to ask if the quantum source coding theorem
may be refined along the lines outlined in the opening paragraph
of section \ref{sect1} (and formulated precisely below).

Since our compression schemes are required to operate with
arbitrarily high fidelity (i.e. reproduce the source states
arbitrarily well as $\epsilon \rightarrow 0$) the question of
whether part of the quantum information of the source may be
represented in classical terms, may be alternatively phrased as
the question of whether it is possible to reversibly extract
classical information from a quantum source in such a way that the
residual quantum information content is reduced. This question has
already been raised in \cite{oldrevext} and in \cite{Winter99} (at
the end of chapter 1).

Consider an encoding operation $E_n$ which encodes
$\ket{\sigma_I}$ into two registers $A$ and $B$ where $A$ holds
the classical part and $B$ holds the quantum part of the encoding.
Let $\{ \ket{j} \}$ be the classical orthonormal basis of $A$. The
most general allowable classical state in $A$ is a probability
distribution over $j$ values so the most general form of the
encoded state may be written \beq \label{state} E_n \ket{\sigma_I}
= \sum_j c_j^I \proj{j} \otimes \omega_j^I \eeq where $\omega_j^I$
are some (generally mixed) states of the subsystem $B$. Here
$c_j^I = p(j|I)$ is the probability of having $j$ in $A$ given
that we are encoding the $I^{\rm th}$ input string. An encoding
operation of this type can be physically interpreted as the action
of an (incomplete) quantum measurement on $\ket{\sigma_I}$. In
this case $j$ is the measurement outcome and $\omega_j^I$ is the
post measurement state (after possible further processing in a way
that can depend on the value of $j$).

>From $p_I$ and $p_j = \sum_I p(j|I)p_I$ we have $p(I|j)= p(j|I)p_I
/p_j$ and for each fixed value of $j$ we get the ensemble \beq
\label{ensj} \ce_j = \{ \omega_j^I ; p(I|j) \}. \eeq Let ${\rm
supp}_j$ be the least number of qubits/signal required to support
the states in $\ce_j$. Then the quantum resource of the encoding
is defined to be \beq \label{avesupp} \overline{\rm supp} = \sum_j
p_j \,\, {\rm supp}_j \eeq We will say that a source $\ce$ may be
compressed to $\alpha$ qubits/signal plus auxiliary classical
storage if for all $\epsilon >0$ there is an $n_0$ such that for
all $n >n_0$ we have an encoding-decoding scheme $(E_n , D_n )$
with fidelity $1-\epsilon$ and $\overline{\rm supp} =\alpha$.

In terms of the above notions the main result of \cite{oldrevext} may
be stated as follows.

\begin{proposition} (\cite{oldrevext}) Let $\ce$ be any irreducible source of
pure states. Let $S$ be the von Neumann entropy of $\ce$ and let
$E_n$ be any encoding scheme for blocks of length $n$ from $\ce$
having a classical and quantum part as in eq. (\ref{state}) above.
Suppose further that\\ (a) the states $\omega_j^I$ in the encoding
are all pure states,
\\ (b) the coding scheme works with fidelity 1 i.e. the states
$\ket{\sigma_I}$ may be {\em perfectly} reconstituted from their
encoded versions.\\ Then the von Neumann entropy of each ensemble
$\ce_j$ is $nS$.\end{proposition}

Hence under the assumptions (a) and (b) it is impossible to reduce
the quantum resource of the encoding below the Schumacher limit $S$
qubits/signal of the original source, by any procedure that extracts
classical information, since $S$ is also the Schumacher limit per
input signal for each $\ce_j$.

The restrictions (a) and (b) are in fact very severe. In particular a
requirement of perfect fidelity (as in (b)) would rule out many basic
theorems of information theory. The compression of classical
information given by Shannon's source coding theorem and the
compression given by the quantum source coding theorem, for example,
would both be impossible. Thus it is of great interest to require
only the weaker condition of asymptotically perfect fidelity i.e.
fidelity of $1-\epsilon$ for all $\epsilon
>0$ where decreasing $\epsilon$ will generally involve working
with increased block lengths $n$. This question was raised in
\cite{oldrevext} but left open.

We now consider the most general situation where both restrictions
(a) and (b) are lifted. Our main results are given in theorems
\ref{3} and \ref{5} below.

\begin{theorem} \label{3}
 Let $\ce = \{ \ket{\sigma_i} ; p_i \}$ be an irreducible
source of pure states with von Neumann entropy $S$ and suppose
that $\alpha \neq S$. Then $\ce$ may be compressed to $\alpha$
qubits per signal plus auxiliary classical storage if and only if
$\alpha>S$.
\end{theorem}

Thus no part of the quantum information content of $\ce$ may be
represented in classical terms if $\ce$ is irreducible. However
this is no longer true if the source is reducible as shown by the
following example. \\ {\bf Example} Consider a reducible ensemble
$\ce = a_1 \ce_1 \cup a_2 \ce_2$ where $\{a_1 ,a_2 \}$ is a
probability distribution and $\ce_i$ are irreducible, supported in
orthogonal subspaces $E_i$ respectively. If $\rho_i$ is the
density matrix of $\ce_i$ then the density matrix $\rho$ of $\ce$
has a block diagonal form \[ \rho = a_1 \rho_1 \oplus a_2 \rho_2
\] so the Schumacher limit of $\ce$ is \[ S(\rho ) = H(a_1 ,a_2 )
+ a_1 S(\rho_1 ) + a_2 S(\rho_2 ) \] where $H(a_1 ,a_2 )$ is the
Shannon entropy of $\{ a_1 ,a_2 \}$.

We have the following encoding scheme for $\ce$: in a long string,
measure each signal (without disturbance) to determine whether it
is in $E_1$ or $E_2$. This provides $H(a_1 ,a_2)$ bits/signal of
classical information. For each value of $i$, Schumacher compress
the signals lying in $E_i$ to $S(\rho_i )$ qubits/signal. The
total quantum resource on average\footnote{To make this statement
precise we need to invoke properties of typical sequences as given
in the proof of theorem \ref{5} later.} is $a_1 S(\rho_1 )+a_2
S(\rho_2 )$ qubits/signal which is less than $S(\rho )$ by the
amount of the classical information extracted. Clearly the
original string may be reconstituted with arbitrarily high
fidelity (for suitably large block lengths) from the classical and
quantum parts of the encoding. Thus if the original ensemble $\ce$
is reducible, it is always possible to convert part of its quantum
information into classical information. In theorem \ref{5} we will
see that the above scheme is actually optimal for providing the
minimal quantum resources in any classical-quantum compression
scheme for a reducible ensemble. \, $\qed$

\begin{theorem} \label{5} Let $\ce = \bigcup_{l=1}^L a_l \ce_l$ be any reducible
ensemble with von Neumann entropy $S$ where $\ce_l$ are
irreducible subensembles supported in orthogonal subspaces. Let
$S_l$ be the von Neumann entropy of $\ce_l$ and suppose that
$\alpha \neq \sum_l a_l S_l $. Then $\ce$ may be compressed to
$\alpha$ qubits per signal plus auxiliary classical storage if and
only if \[ \alpha > \sum_l a_l S_l = S-H(a_1 , \ldots , a_L ).
 \] \end{theorem}

Note that if we do not require the residual quantum resource of the
encoding to be smaller than that of the original source then
reversible extraction of classical information (even with perfect
fidelity) is always possible. Indeed as described in \cite{oldrevext}
the process of quantum teleportation may be interpreted as a scheme
for encoding a quantum source into classical and quantum information
with perfect fidelity in decoding, but the associated quantum
resource of the encoding is not less than that of the original
source. Also one may consider the trivial encoding of retaining the
input string untouched and merely attaching independent classical
information, which is discarded in the decoding.

We will approach the proofs of theorem \ref{3} and theorem \ref{5}
through a series of lemmas. Firstly lemma \ref{9} below will
relate the quantum resource of any encoding-decoding scheme to the
amount of mutual information per signal, between the classical
part of the encoding and the identity of the input string. Then we
will use lemmas \ref{10} and \ref{11} to show that this mutual
information must tend to zero for irreducible sources, as the
fidelity of the scheme tends to unity, and we will also
characterise its limiting value for reducible sources.

\section{Mutual information of the classical extraction}
\label{sect5}

\begin{lemma} \label{9} Let $\ce$ be a source with von Neumann entropy $S$. Let
$(E_n , D_n )$ be any encoding-decoding scheme for blocks of
length $n$ with average fidelity $1-\epsilon$. Suppose that the
encoded states have a classical and quantum part as in eq.
(\ref{state}). Let $\overline{\rm supp}$ be the quantum resource
of the encoding (as in eq. (\ref{avesupp})) and let $\ci (I:J)$
denote the mutual information between the input string $I$ and the
classical data $j$ i.e. the mutual information of the probability
distribution $p(I \& j) =p_I \, c^I_j $. Then \beq \label{lemma3}
\overline{\rm supp} +\frac{\ci (I:J)}{n} \geq S -f(\epsilon ) \eeq
where $f(\epsilon ) $ is a function satisfying
\[ f(\epsilon ) \rightarrow 0 \hspace{5mm} \mbox{ as } \hspace{5mm}
\epsilon \rightarrow 0.\] \end{lemma} {\bf Remark} We will
actually prove a slightly stronger statement. Let $(E,D)$ be an
encoding-decoding scheme for a source $\ce_{in}$ in dimension $d$
with von Neumann entropy $S_{in}$. Let $\overline{\rm Supp}$ be
the average number of qubits needed to support the intermediate
ensembles $\ce_j$ and let $I$ denote the identity of the input
state. Then \beq \label{lemma3prime} \overline{\rm Supp} + \ci
(I:J) \geq S_{in} -f(\epsilon ) \log d \eeq In lemma \ref{9},
$\ce_{in}$ has the form $\ce^{\otimes n}$ (i.e. $n$-strings from
$\ce$) so $S_{in}=nS$, $\overline{\rm Supp} = n\overline{\rm
supp}$ and $d=k^n$ where $k$ is the dimension of the single signal
space.

{\bf Proof} Let us write the encoded states as \beq \tau_I = E_n
\ket{\sigma_I} = \sum_j p(j|I) \proj{j}\otimes \omega_j^I . \eeq
These states all have a block diagonal form with blocks labelled
by $j$ containing $p(j|I)\omega_j^I$. Consider the ensemble
$\ce_{enc} = \{ \tau_I ; p_I \}$. The average state is
$\overline{\tau} = \sum_I p_I \tau_I$ and the Holevo quantity of
$\ce_{enc}$ is \[ \chi_{enc}= S(\overline{\tau}) - \sum_I p_I
S(\tau_I ). \] If we fix on any single value of $j$  we have the
ensemble $\ce_j = \{ \omega_j^I ; p(I|j) \}$. Let $\chi_j$ be the
Holevo quantity of $\ce_j$. Using the block diagonal form of
$\ce_{enc}$ a straightforward rearrangement of the formula for
$\chi$ gives \beq \label{xenc} \chi_{enc} = \sum_j p_j \chi_j +
\ci (I:J). \eeq For any ensemble the Holevo quantity satisfies
$\chi \leq \log d$ where $d$ is the dimension of the space of
states for the ensemble. Hence \beq \label{supx} \chi_j \leq n\,\,
{\rm supp}_j \hspace{1cm} \mbox{for all $j$.} \eeq and from eqs.
(\ref{supx}) and (\ref{xenc}) \beq \label{ineq1} n\,\,
\overline{\rm supp} +\ci (I:J) \geq \chi_{enc}. \eeq

Now consider the decoding stage. We have \[ D_n (\tau_I ) =
\tilde{\sigma}_I \] with average fidelity
\[ \overline{F} = \sum_I p_I F(\ket{\sigma_I}, \tilde{\sigma}_I )
= 1-\epsilon. \] Let $\chi_{dec}$ be the Holevo quantity of the
decoded ensemble $\ce_{dec} = \{ \tilde{\sigma}_I ; p_I \}$ and let
$\chi_{in} = nS$ be the Holevo quantity of the input ensemble $\ce =
\{ \ket{\sigma_I}; p_I \}$. We will use the result, proved in
appendix A, that high fidelity ensembles have close $\chi$'s. More
precisely, since $\ce_{in}$ and $\ce_{dec}$ (supported in dimension
$d=k^n$ where $k$ is the dimension of the single signal space) have
fidelity $1-\epsilon$ we can say \beq |\chi_{in}-\chi_{dec}| \leq
4(\sqrt{\epsilon} \log(k^n ) -\sqrt{\epsilon} \log
(2\sqrt{\epsilon})) \eeq so \beq \label{ineq2}  \chi_{dec} \geq nS -
4n\sqrt{\epsilon} \log k + 4\sqrt{\epsilon}\log (2\sqrt{\epsilon})
\eeq Now the decoding operation $D_n$ is a CPTP map and by the
Uhlmann-Lindblad monotonicity theorem \cite{Lindblad73a,Uhlmann77a}
the Holevo quantity is non-increasing under any CPTP map. Thus
$\chi_{enc} \geq \chi_{dec}$ and eqs. (\ref{ineq1}) and (\ref{ineq2})
give
\[ n\,\, \overline{\rm supp} +\ci (I:J) \geq nS - 4n \sqrt{\epsilon}
\log k + 4\sqrt{\epsilon}\log (2\sqrt{\epsilon}) \] so \[
\overline{\rm supp} + \frac{\ci (I:J)}{n} \geq S - f(\epsilon ) \]
where $f(\epsilon ) \rightarrow 0$ as $\epsilon \rightarrow 0$, as
required.\, $\qed$

\section{An information--disturbance relation} \label{sect6}

To complete the proof of theorem \ref{3} we will argue that $\ci
(I:J)/n$ must also tend to zero as $\epsilon$ tends to zero. The
intuitive reason is the following. After encoding and decoding
(thinking of $\epsilon$ as being very small) the states
$\tilde{\sigma}_I$ reproduce the states $\ket{\sigma_I}$ with high
fidelity. But the classical data $j$ can be assumed to remain
after the process since it may be copied at the encoding stage
into another register which is not affected by the decoding
operation. Now it is a general heuristic principle in quantum
physics that one cannot obtain information about a fixed source of
non-orthogonal states without disturbing them and furthermore
there should be a tradeoff between the amount of disturbance and
the amount of information gained. The fundamental role of
information-disturbance tradeoffs in quantum measurement theory
has been emphasised by C. A. Fuchs \cite{fuchs}. However we need a
more refined version of this principle as our input source
generally varies (because of increasing block lengths) as the
disturbance $\epsilon$ tends to zero. Nevertheless we will show
that $\ci (I:J )/n$, the information gained per signal, goes to
zero as the fidelity approaches 1.

Indeed in the limiting case of perfect fidelity (i.e. where
$\tilde{\sigma}_I = \proj{\sigma_I}$ ) it is not difficult to show
that $\ci (I: J)$ must be exactly zero (cf \cite{bbm}). The proof
is as follows: Any CPTP map may be represented as a unitary
operation acting on the input together with an ancilla (in some
standard initial state $\ket{0}$) followed by tracing over a
subsystem of the output. Thus the encoding and decoding operation
on $\ket{\sigma_I}$ may be represented as a unitary operation $U$
on $\ket{\sigma_I}_A\ket{0}_B$ in registers A and B where B is the
ancilla, yielding a pure state \[ U (\ket{\sigma_I}_A \ket{0}_B )
= \ket{\lambda_I}_{AB}\] where $\tilde{\sigma}_I = \tr_B
\proj{\lambda}$ and the classical data is obtained from a
subsystem of $\tr_A \proj{\lambda}$. If the $\ket{\sigma_I}$'s are
reproduced with perfect fidelity we must have \[
\ket{\lambda}_{AB} = \ket{\sigma_I}_A \ket{\psi_I}_B \] for some
pure states $\ket{\psi_I}$. But then from the unitarity of $U$
\[ \inner{\sigma_I}{\sigma_K} \inner{0}{0} =
\inner{\sigma_I}{\sigma_K} \inner{\psi_I}{\psi_K}. \] Hence if there
is a chain from $\ket{\sigma_I}$ to $\ket{\sigma_K}$ we must have
$\inner{\psi_I}{\psi_K}=1$ i.e. $\ket{\psi_I}=\ket{\psi_K}$. If $\ce$
is irreducible then this is true for all $I$ and $K$ so no
measurement on register B can yield any information about the
identity of $I$. In particular $\ci (I:J)$ must be zero.

In lemmas \ref{10} and \ref{11} below we will generalise the above
argument to the scenario of arbitrarily high (but not perfect)
fidelity, showing that $\ci (I:J)/n \rightarrow 0$ as $\epsilon
\rightarrow 0$.

\begin{lemma} \label{10}
 Suppose that $\ce = \{ \ket{\sigma_i};p_i \}$ is an
irreducible source with $K$ states. Suppose that the states
$\ket{\sigma_i}$ are provided in a register A with state space
$\cb_{\alpha_1}$ and let register B be an ancilla with state space
$\cb_{\alpha_2}$. We will refer to B as the environment. Let \beqa
\label{gamma} \Gamma : \cb_{\alpha_1}\otimes \cb_{\alpha_2}
\rightarrow \cb_{\alpha_1}\otimes \cb_{\alpha_2} \\ \Gamma
\ket{\sigma_i}_{A} \ket{0}_B = \ket{\xi_i}_{AB} \eeqa be a unitary
map such that \beq \sum_i p_i F\left( \ket{\sigma_i}, \tr_B
\proj{\xi_i}\right) = 1-\epsilon \eeq Let  $\{ \rho_i = \tr_A
\proj{\xi_i}; p_i \}$ be the environment ensemble and let \[ \chi
= S(\sum_i p_i \rho_i ) -\sum_i p_i S(\rho_i ) \] be the Holevo
quantity of the environment. Then if $\ce$ is kept fixed but
$\epsilon$, $\Gamma$ and $\alpha_2$ are allowed to vary, we have
$\chi \leq f(\epsilon )$ where the function $f$ satisfies
$f(\epsilon ) \rightarrow 0$ as $\epsilon \rightarrow 0$. In fact
we may take $f(\epsilon ) = A\sqrt{\epsilon} +B\sqrt{\epsilon}
\log \sqrt{\epsilon}$ where $A$ and $B$ are constants. \end{lemma}
{\bf Remark} We are thinking here of $\Gamma$ as being a unitary
extension of a CPTP coding-decoding map $D_n E_n$ with high
fidelity $1-\epsilon$. Note that any coding--decoding scheme for
any source may be assumed to be of the form $\Gamma$ in eq.
(\ref{gamma}) where register A contains the final decoded state
and the environment B may, without loss of generality, be assumed
to retain a copy of the classical part of the encoding (since it
may be copied after encoding and the copy kept intact during
decoding). Thus $\tr_B \proj{\xi_i}$ is the decoded version of the
input $\ket{\sigma_i}$ and by Holevo's theorem \cite{Holevo73}
$\chi$ of the environment is an upper bound for the amount of
information about $i$ that may be obtained by any measurement on
B. Thus the lemma states that any such information must approach
zero as the average disturbance to the ensemble tends to zero.

The proof of lemma \ref{10} is given in appendix B.

Note that in lemma \ref{10} the source is kept constant as
$\epsilon$ varies: there is no notion of increasing block length
as $\epsilon \rightarrow 0$. For our desired application in
theorem \ref{3} the ensemble $\ce$ varies as $\epsilon \rightarrow
0$ since the block length generally increases. The proof of lemma
\ref{10} is not directly applicable in this situation (as the
parameters $K$ and $\zeta$ also increase with block length) and
this extra complication is dealt with in lemma \ref{11} below. As
a preliminary result we have:

\begin{lemma} {\bf (Markov Lemma)} Let $\{ X_i ; p_i \}$ be any random
variable with $0\leq X_i \leq 1$ and mean $\overline{X} >
1-\epsilon$. Then for any $A$ we have $Prob(X_i < 1-A\epsilon )
<\frac{1}{A}$. In particular
\[ Prob(X_i < 1-\sqrt{\epsilon}) <\sqrt{\epsilon}. \] \end{lemma}

{\bf Proof} If  $Prob(X_i < 1-A\epsilon ) =\alpha$  we get
\[ 1-\epsilon < \sum p_i X_i = \sum_{X_i < 1-A\epsilon} p_i X_i +
\sum_{X_i \geq 1-A\epsilon} p_i X_i \leq (1-A\epsilon ) \alpha+
(1-\alpha) 1 = 1-A\alpha\epsilon \] Hence $A\alpha <1$. $\qed$.

\begin{lemma} \label{11}
 Suppose we have a sequence $\{ \epsilon_m >0 \}$ with
$\epsilon_m \rightarrow 0$ and let $n(m)$ be any (generally
unbounded) function of $m$. Suppose also that for each $m$ we
have:
\\(i) a source $\ce^{(m)} = \ce_1^{(m)}\otimes \ldots \otimes
\ce_{n(m)}^{(m)}$ where each $\ce^{(m)}_i$ is an irreducible
source on a state space of at most $k$ dimensions with at most $K$
signal states. \\(ii) An encoding-decoding scheme $(E^{(m)},
D^{(m)})$ on $\ce^{(m)}$ with average fidelity $1-\epsilon_m$,
leaving the environment in a state $\rho_I^{(m)}$ for input state
labelled by $I = i_1\ldots i_{n(m)}$.\\ Then
$\chi(\{\rho_I^{(m)};p_I^{(m)} \})/n(m) < g(\epsilon_m)$ where $g$
is a function satisfying $g(\epsilon ) \rightarrow 0$ as $\epsilon
\rightarrow 0$. Hence the amount of information per position tends
to zero as the fidelity tends to 1, for arbitrarily changing block
lengths in the schemes. \end{lemma} {\bf Remark} In our
application below of lemma \ref{11} to an irreducible ensemble
$\ce$, $\ce^{(m)}$ will be the ensemble of strings of length
$n(m)$ from $\ce$ so $\ce_i^{(m)} = \ce$ for all $i$ and $m$. For
reducible ensembles $\ce$ however, it will be necessary to
consider the irreducible parts of the ensemble of $n(m)$-strings
so that each $\ce_i^{(m)}$ will range over the various irreducible
subensembles of $\ce$ and we will need lemma \ref{11} in its full
generality.

{\bf Proof of lemma \ref{11}} Let us fix attention on any one of
the schemes labelled by $m$ and omit reference to the value of $m$
in all labels, for notational clarity. We write $I_{\neq l}$ for
the input string with the $l^{\rm th}$ position deleted.

Let $\tau_{i_1 \ldots i_n}$ be the decoded output for the input
string $\ket{\sigma_{i_1 \ldots i_n}}=\ket{\sigma_{i_1}} \ldots
\ket{\sigma_{i_n}} \in \ce_1 \otimes \ldots \ce_n $ and write
\[ F_I \equiv  F_{i_1 \ldots i_n} = \bra{\sigma_{i_1 \ldots i_n}}
\tau_{i_1 \ldots i_n} \ket{\sigma_{i_1 \ldots i_n}}. \] Let $p_I
\equiv p_{i_1 \ldots i_n} = p_{i_1}^{(1)} \ldots p_{i_n}^{(n)}$
where $p_i^{(k)}$ is the probability of $\ket{\sigma_i}$ in the
ensemble $\ce_k$. For notational clarity we will henceforth omit
the superscript $(k)$ on the probabilities. Then \beq \label{fids}
F = \sum_{i_1 \ldots i_n} p_{i_1 \ldots i_n} F_{i_1 \ldots i_n}
=1-\epsilon. \eeq Let $\{ \rho_I ; p_I \}$ be the ensemble of
final environment states of the coding--decoding scheme. We will
use the following inequality, proved in appendix C, for the Holevo
quantity of the environment:
\begin{equation}\label{chichi}
\frac{\chi( \{\rho_I;p_I \} )}{n} \leq  \max_k \sum_{I_{\neq k}}
p_{I_{\neq k}}
    \chi_{I_{\neq k}},
\end{equation}
where \[ \chi_{I_{\neq k}} =S( \sum_{i_k} p_{i_k} \rho_I )
        - \sum_{i_k} p_{i_k} S( \rho_I ) \]
and we will argue that each term on the RHS of eq. (\ref{chichi})
tends to zero with $\epsilon$.

Consider $k=1$ (all others are similar). For each fixed choice of
$I_{\neq 1} = i_2 \ldots i_n$ we extend $\ket{\sigma_i}$ in the
$I_1$ slot by $\ket{\sigma_{i_2 \ldots i_n}}$, apply the operation
$DE$, and look at the $I_1$ slot of the output. This is a
coding/decoding of $\ket{\sigma_i}$ (i.e. length 1 string from
$\ce_1$) with output $\tau_i = \tr_{i_2 \ldots i_n} \tau_{i i_2
\ldots i_n}$. Furthermore $\chi_{I_{\neq 1}}$ is the Holevo
quantity of the environment after this coding--decoding of
$\ce_1$. The fidelity is \beq
\begin{array}{rcl} F^{(i_2 \ldots i_n )} & = & \sum_i p_i
\bra{\sigma_i} \tau_i \ket{\sigma_i} \\
 & = &  \sum_i p_i
\bra{\sigma_i} \tr_{i_2 \ldots i_n} \tau_{i i_2 \ldots i_n}
\ket{\sigma_i}
\\
 & \geq & \sum_i p_i
\bra{\sigma_{ii_2 \ldots i_n}}  \tau_{i i_2 \ldots i_n}
\ket{\sigma_{ii_2 \ldots i_n}} = \sum_i p_i F_{i i_2 \ldots i_n}
\end{array} \eeq (Here the last inequality arises since we can extend
$\ket{\sigma_{i_2 \ldots i_n}}$ to an orthonormal basis of the
$I_2 \ldots I_n$ slots to perform the partial trace.)

Next we apply the Markov lemma to the random variable  \[ \{ X_{i_2
\ldots i_n} \equiv F^{(i_2 \ldots i_n )} = \sum_i p_i F_{ii_2 \ldots
i_n}\, ; \, p_{i_2 \ldots i_n} \} \] (noting that eq. (\ref{fids})
gives $\overline{X} > 1-\epsilon$) to conclude:
\[ \sum_i p_i F_{ii_2 \ldots i_n} < 1-\sqrt{\epsilon} \hspace{5mm}
\mbox{with probability $<\sqrt{\epsilon}$.} \] Divide strings $i_2
\ldots i_n$ (taken with probabilities $p_{i_2 \ldots i_n}$) into
\[ S_{good} = \{ i_2 \ldots i_n | \sum_i p_i F_{ii_2 \ldots i_n} >
1-\sqrt{\epsilon} \} \hspace{5mm} \mbox{with total probability
$>1-\sqrt{\epsilon}$}
\]
\[ S_{bad} = \{ i_2 \ldots i_n | \sum_i p_i F_{ii_2 \ldots i_n}<
1-\sqrt{\epsilon} \} \hspace{5mm} \mbox{with total probability
$<\sqrt{\epsilon}$}
\] i.e. $S_{good}$ are those extensions of $I_1$ which retain high
fidelity for reproducing the first slot after coding/decoding of the
extension. Now
\[ \sum_{I_{\neq 1}} p_{I_{\neq 1}} \chi_{I_{\neq 1}}
 = \sum_{good} (same) +
\sum_{bad} (same). \] For good sequences, lemma \ref{10}  then
gives $\chi_{I_{\neq 1}} \leq f(\sqrt{\epsilon} )$ (as fidelity of
the coding/decoding is $>1-\sqrt{\epsilon}$) where $f$ is a
function satisfying $f(x) \rightarrow 0$ as $x\rightarrow 0$. For
bad sequences we always have $\chi_{I_{\neq 1}} \leq \log k$,
where $k$ is the dimension of the one-signal space. This is
because for each value of $I_{\neq 1}$ the ensemble $\{ \rho_{i_1
i_2 \ldots i_n }; p_{i_1}\}_{i_1}$ is obtained by a CPTP map from
$\{ \ket{\sigma_{i_1}}; p_{i_1} \}$ so from the Uhlmann-Lindblad
monotonicity theorem we get $\chi_{I_{\neq 1}} \leq \chi (\ce_1 )
\leq \log k$.

Hence from the weights of the good and bad sets we get
\[ \chi_{I_{\neq 1}} \leq (1-\sqrt{\epsilon})
f(\sqrt{\epsilon}) +\sqrt{\epsilon} \log k \equiv g(\epsilon )
\] where clearly $g(\epsilon ) \rightarrow 0$ as
$\epsilon \rightarrow 0$. $\qed$.

\section{Completing the main proofs} \label{sect7}

Finally we assemble our lemmas to provide proofs of theorems
\ref{3} and \ref{5}.

{\bf Proof of theorem \ref{3}} Suppose that $\ce$ can be
compressed to $\alpha$ qubits/signal plus auxiliary classical
storage. Then for each $\epsilon >0$ and all sufficiently large
$n$ there is an encoding-decoding scheme which, by lemma \ref{9}
satisfies \beq \label{pf3} \alpha + \frac{\ci (I:J)}{n} \geq
S-f(\epsilon ). \eeq Here $f(\epsilon ) \rightarrow 0$ as
$\epsilon \rightarrow 0$ and by lemma \ref{11} (with $\ce_i = \ce$
for all $i$) we have $\ci(I:J)/n \rightarrow 0$ as $\epsilon
\rightarrow 0$ too. But eq. (\ref{pf3}) holds for all $\epsilon
>0$ so if $\alpha \neq S$ we must have $\alpha >S$.

Conversely if $\alpha >S$ then $\ce$ may be compressed to $\alpha$
qubits/signal using just standard Schumacher compression (and no
auxiliary classical storage). \,$\qed$

For the proof of theorem \ref{5} we will use the following
standard result:

\begin{lemma} {\bf (Lemma of typical sequences)}\cite{covth}
Let $\cp = \{
p_1 , \ldots ,p_L \}$ be any probability distribution and consider
sequences $i_1 \ldots i_n$ of the symbols $1, \ldots ,L$ with
probabilities $p_{i_1} \ldots p_{i_n}$. Let $n(i)$ be the number
of times that the symbol $i$ occurs in the sequence. For any
$\epsilon
>0$ and $n$  let $S_n (\epsilon ) = \{ i_1 \ldots i_n : | n(i) -np_i | <
L\frac{\sqrt{n}}{\sqrt{\epsilon}}  \}$. Then for any $\epsilon
>0$ there is an $n_0$ such that for all $n>n_0$ the total probability
of $S_n (\epsilon )$ is greater than $1-\epsilon$. For such
sufficiently large $n$, $S_n (\epsilon )$ is called a set of
$\epsilon$-typical sequences. \end{lemma}

Thus a sequence is typical if the frequency of occurrence of each
symbol $i$ in it is approximately equal to the prior probability
$p_i$.

{\bf Proof of theorem \ref{5}} We have a signal ensemble $\ce
=\bigcup_{l=1}^L a_l \ce_l$ where $\ce_l$ are irreducible
ensembles supported in orthogonal subspaces. Let $\ca$ denote the
probability distribution $\{a_1 ,\ldots ,a_L \}$.

Suppose that $\ce$ can be compressed into $\alpha$ qubits per signal
plus auxiliary classical storage. Then for each $\epsilon >0$ and all
sufficiently large $n$ there is an encoding-decoding scheme for
$n$-strings with fidelity $F>1-\epsilon$ and $\overline{\rm supp}
=\alpha$.

The source of all $n$-strings from $\ce$ decomposes into irreducible
parts: \[ \ce^{\otimes n} = \bigcup_{l_1 \ldots l_n} a_{l_1} \ldots
a_{l_n} \ce_{l_1} \otimes \ldots \otimes \ce_{l_n} \] and the
fidelity may be expressed as an average:
\[ F = \sum_{l_1 \ldots l_n}a_{l_1} \ldots
a_{l_n} F_{l_1 \ldots l_n}>1-\epsilon \] where $F_{l_1 \ldots
l_n}$ is the fidelity of the scheme when restricted to $ \ce_{l_1}
\otimes \ldots \otimes \ce_{l_n}$. We will apply lemmas \ref{9}
and \ref{11} to these irreducible parts. For any sequence $l_1
\ldots l_n$ let $n(l)$ denote the number of times that the symbol
$l=1,\ldots L$ occurs. Then the von Neumann entropy of $ \ce_{l_1}
\otimes \ldots \otimes \ce_{l_n}$ is $\sum_l n(l) S_l$.

By applying the Markov lemma to the random variable $\{  F_{l_1
\ldots l_n};a_{l_1} \ldots a_{l_n} \}$ we obtain a set of sequences
\[ S_{good} = \{ l_1 \ldots l_n :  F_{l_1 \ldots
l_n}>1-\sqrt{\epsilon} \} \hspace{4mm} \mbox{with total probability
$>1-\sqrt{\epsilon}$}
\]
By selecting the $\epsilon$-typical subset of these we get
\[  S_{good,typ} = \{ l_1 \ldots l_n :  F_{l_1 \ldots
l_n}>1-\sqrt{\epsilon} \hspace{3mm} \mbox{and $l_1 \ldots l_n$ is
$\epsilon$-typical} \} \] with total probability
$>1-2\sqrt{\epsilon}$.

Consider the compression scheme acting on the irreducible
component $ \ce_{l_1} \otimes \ldots \otimes \ce_{l_n}$. For any
one of the good sequences lemma \ref{9} gives
\[ n\alpha +\ci (I_{l_1 \ldots l_n}:J) \geq \sum_l n(l)S_l
-n f(\sqrt{\epsilon} ) \] where $\ci (I_{l_1 \ldots l_n}:J)$ is
the mutual information for the source of restricted $n$-strings
and $f(x)\rightarrow 0$ as $x\rightarrow 0$. Furthermore lemma
\ref{11} gives $ \ci (I_{l_1 \ldots l_n}:J)/n <g(\sqrt{\epsilon})$
where $g(x)\rightarrow 0$ as $x\rightarrow 0$ too. Hence
\[ \alpha \geq \sum_l \frac{n(l)}{n}S_l
-f(\sqrt{\epsilon})-g(\sqrt{\epsilon}) \] If our chosen good sequence
is also typical then
\[ a_l -\frac{L}{\sqrt{\epsilon}\sqrt{n}} <  \frac{n(l)}{n} <
a_l +\frac{L}{\sqrt{\epsilon}\sqrt{n}} \] so for each fixed
$\epsilon$, $ \frac{n(l)}{n} \rightarrow a_l$ as $n\rightarrow
\infty$. Thus
\[ \alpha \geq \sum_l a_l S_l -f(\sqrt{\epsilon}) -g(\sqrt{\epsilon})
\]
and finally letting $\epsilon \rightarrow 0$ we get $\alpha \geq \sum
a_l S_l$ as required.

Conversely to see that the bound is tight let $\epsilon>0$ and
$\delta>0$ be any chosen values and let $\alpha =\sum_l a_l S_l
+\delta$. For all sufficiently large $n$ we encode an $n$-string from
$\ce^{\otimes n}$  as follows. We first measure each signal (without
disturbance) to determine which sub-ensemble $\ce_l$ it belongs to,
giving a string $l_1 \ldots l_n$ drawn from $\ca^n$. If the sequence
$l_1 \ldots l_n$ is $\epsilon$-typical for $\ca$ (so each value  $l$
occurs between $a_l n \pm L\sqrt{n}/\sqrt{\epsilon}$ times) we
perform Schumacher compression to $na_l S_l + O(\sqrt{n})$ qubits for
each value of $l$, giving $\sum a_l S_l +O(\frac{1}{\sqrt{n}})$
qubits/signal overall. If the string is atypical we generate an
arbitrary fixed state of $n\alpha$ qubits. By the dominating weight
of typical sequences and the asymptotic fidelity of Schumacher
compression, this scheme clearly has fidelity $1-O(\epsilon )$ and
for all sufficiently large $n$, the quantum resource will be less
than $\alpha = \sum a_l S_l +\delta$ qubits/signal. $\qed$

\section{Concluding remarks}\label{sect8}

We have shown that no part of the quantum information of an
irreducible source may be replaced by classical information if
arbitrarily long strings are to be reconstitutable with
asymptotically high fidelity $1-\epsilon$ for all $\epsilon >0$. Also
for reducible sources we have characterised the maximum possible
amount of classical information that can be reversibly extracted from
the source under the above conditions.

To obtain these results we first proved some information--disturbance
relations. Let $\ce$ be an irreducible source. If some mutual
information $\ci$ is obtained about the identity of the state from a
single emission from $\ce$ by any physical process, leaving the state
intact with average fidelity $1-\epsilon$ then we showed that
$\ci\rightarrow 0$ as $\epsilon \rightarrow 0$. For strings of length
$n$ from $\ce$ we considered a sequence of encoding-decoding schemes
(labelled by $m=1,2, \ldots $) with asymptotically perfect fidelity
($1-\epsilon_m \rightarrow 1$) for which the string length $n(m)$ may
vary arbitrarily (e.g. grow unboundedly) with $m$. In this case we
showed that the mutual information {\em per letter} $\ci_m /n(m)$
provided by the $m^{\rm th}$ scheme, must tend to zero as the
fidelity tends to 1. This was sufficient for our purposes but raises
the interesting question of a possibly stronger result: does $\ci_m$
itself necessarily tend to zero too, as $\epsilon_m \rightarrow 0$,
or can $\ci_m$ remain nonzero under these conditions (while $\ci_m /n
\rightarrow 0$)? Consider for example an irreducible source $\ce = \{
\ket{\sigma_0}, \ket{\sigma_1};p_0 ,p_1 \}$ of two non-orthogonal
states. Is it possible to have a sequence of encoding-decoding
schemes $(E_n ,D_n )$ for strings of increasing length $n$, such that
the $n^{\rm th}$ scheme has average fidelity $1-\epsilon_n$ with
$\epsilon_n \rightarrow 0$ and the $n^{\rm th}$ scheme provides
$\ci_n = 1$ bit of information about the identity of the string? Here
$\ci_n /n \rightarrow 0$ so our result is not contradicted yet a
nonzero amount of information about the whole string is obtained with
vanishing disturbance to the states. This question remains open.

\bigskip

\noindent {\Large\bf Acknowledgements}

RJ is supported by the U.K. Engineering and Physical Sciences
Research Council and PH is supported by the Rhodes Trust. AW is
supported by SFB 343 ``Diskrete Strukturen in der Mathematik'' of the
Deutsche Forschungsgemeinschaft. Part of this work was carried out
during collaborative visits supported by the European Science
Foundation. HB, PH and RJ also acknowledge the support of the
European 5$^{\rm th}$ framework network QAIP IST-1999-11234. \bigskip

\noindent {\Large\bf Appendix A}

\begin{proposition}  Let $\ce_1 = \{ \omega_k ; p_k \}$ and $\ce_2 =
\{ \rho_k ; p_k \}$ be two ensembles on the same space, of
dimension $d$,  with the same prior probabilities. Here $\omega_k$
and $\rho_k$ are generally mixed states. Let $\chi_1$ and $\chi_2$
be their Holevo quantities. Suppose
\[ \overline{F} = \sum_k p_k F(\omega_k , \rho_k ) =
1-\epsilon
\] with $\epsilon \leq 1/16$. Then \beq \label{a1}
|\chi_1 - \chi_2 | \leq
4(\sqrt{\epsilon} \log d -\sqrt{\epsilon} \log (2\sqrt{\epsilon})
\eeq
\end{proposition}

{\bf Proof} In \cite{michalmixed} it is shown that \beq \label{a2}
|\chi_1 -\chi_2 | \leq 2(\epsilon ' \log d -\epsilon ' \log
\epsilon ' ) \eeq where \[ \sum_i p_i ||\omega_i -\rho_i || =
\epsilon ' \leq \frac{1}{2} \] and $||\omega ||$ denotes the trace
norm (i.e. the sum of the absolute values of the eigenvalues) of
$\omega$. This norm is related to our fidelity function by
\cite{Winter99,Fuchs1999a} \[ ||\rho -\omega || \leq 2
\sqrt{1-F(\rho ,\omega )}. \] Let $\epsilon_k = 1-F(\rho_k
,\omega_k )$. Then \beq \label{a3} \epsilon ' = \sum p_k
||\omega_k -\rho_k || \leq 2 \sum p_k \sqrt{\epsilon_k } \leq
2\sqrt{ \sum p_k \epsilon_k } = 2\sqrt{\epsilon}. \eeq Thus if
$\epsilon \leq \frac{1}{16}$ we have $\epsilon ' \leq \frac{1}{2}$
and eqs. (\ref{a2}) and (\ref{a3}) give the required inequality
eq. (\ref{a1}).\, $\qed$

We note that a slightly weaker form of eq. (\ref{a1}) is proved by
different means in \cite{mixed}.\bigskip

\noindent {\Large\bf Appendix B}

{\bf Proof of lemma \ref{10}} Let $\eta$ be the smallest non-zero
overlap $|\inner{\sigma_i}{\sigma_j}|$ of any two signals
$\ket{\sigma_i}$ and $\ket{\sigma_j}$. We have
\[ \Gamma :\cb_{\alpha_1}\otimes \cb_{\alpha_2} \rightarrow
\cb_{\alpha_1}\otimes \cb_{\alpha_2} \] mapping $\ket{\sigma_i}_{A}
\ket{0}_B$ to $\ket{\xi_i}_{AB}$. Write \[ \tilde{\sigma}_i = \tr_B
\proj{\xi_i}. \] The average fidelity is
\[ \overline{F} = \sum_i p_i F(\ket{\sigma_i} , \tilde{\sigma}_i )
= 1-\epsilon . \] For each $i$ let $F(\ket{\sigma_i} ,
\tilde{\sigma}_i ) = 1-\epsilon_i$ so that \beq \sum_i p_i \epsilon_i
=\epsilon \label{aveeps} \eeq Since we have $\ce$ fixed but may think
of $\epsilon$ as varying, we have $\epsilon_i = O(\epsilon )$ for
each $i$.

For each value of $i$ we consider an orthonormal basis of register A
which has $\ket{\sigma_i}$ as its first member:
\[ \{ \ket{\sigma_i}, \ket{\tau_1},\ket{\tau_2}, \ldots \}. \]
Since $\ket{\xi_i}$ has fidelity $1-\epsilon_i$ to be
$\ket{\sigma_i}$ in register A, we can write: \beq\label{XI}
\ket{\xi_i}_{AB} = \sqrt{1-\epsilon_i} \ket{\sigma_i}_A
\ket{\beta_i}_B + \sqrt{\epsilon_i} \ket{\gamma_i}_{AB} \eeq where
the normalised state $\ket{\gamma_i}_{AB}$ has the form
\beq\label{gtau} \ket{\gamma_i} = \sum_{m\geq 1} a_m
\ket{\tau_m}\ket{\delta_m} \eeq and $\ket{\beta_i},\ket{\delta_1},
\ket{\delta_2}, \ldots $ are some normalised states of B (which
generally all vary with $i$). For any other value $k$ of $i$ we have
correspondingly
 \beq\label{XK}
\ket{\xi_k}_{AB} = \sqrt{1-\epsilon_k} \ket{\sigma_k}_A
\ket{\beta_k}_B + \sqrt{\epsilon_k} \ket{\gamma_k}_{AB}. \eeq Our
strategy is the following: thinking of $\epsilon_i$ and $\epsilon_k$
as small we note that the $\ket{\gamma}_{AB}$ terms have small
amplitude and we will now argue that the states $\ket{\beta_i}$ and
$\ket{\beta_k}$ are then also close. Hence the reduced states in
register B for different values of $i$ will be almost independent of
$i$ and hence will have very low $\chi$. Correspondingly any
measurement on B can provide only very little information about the
identity of $i$. For notational clarity we will sometimes write the
product state $\ket{\alpha}\ket{\beta}$ of registers AB as
$\ket{\alpha \beta}$.

The unitarity of $\Gamma$ with the expressions eqs. (\ref{XI}) and
(\ref{XK}) gives \beqa \inner{\sigma_i}{\sigma_k} \inner {0}{0} =
\inner{\xi_i}{\xi_k} =  \sqrt{1-\epsilon_i}\sqrt{1-\epsilon_k} \,
\inner{\sigma_i}{\sigma_k} \, \inner {\beta_i}{\beta_k}+ \\
\sqrt{1-\epsilon_i}\sqrt{\epsilon_k}\, \bra{\sigma_i \beta_i}
\gamma_k \rangle + \sqrt{1-\epsilon_k}\sqrt{\epsilon_i}\,
\bra{\gamma_i} \sigma_k \beta_k \rangle +
\sqrt{\epsilon_i}\sqrt{\epsilon_k}\, \inner{\gamma_i}{\gamma_k}.
\eeqa Each inner product in the last three terms is a complex number
with modulus at most one. Let us denote them by $a_1 , a_2$ and $a_3$
and write $\mu = \inner{\sigma_i}{\sigma_k}$. Then \beq\label{mess}
\sqrt{1-\epsilon_i}\sqrt{1-\epsilon_k} \inner{\beta_i}{\beta_k} =
1-\frac{( \sqrt{1-\epsilon_i}\sqrt{\epsilon_k}\, a_1 +
\sqrt{1-\epsilon_k}\sqrt{\epsilon_i}\, a_2 +
\sqrt{\epsilon_i}\sqrt{\epsilon_k}\, a_3  )}{\mu}. \eeq Now if
$\epsilon$ is sufficiently small we will have
\[ \left| \frac{\sqrt{\epsilon_i}}{\mu}\right| <
\frac{1}{3}\hspace{5mm} \mbox{ and } \hspace{5mm}  \left|
\frac{\sqrt{\epsilon_k}}{\mu}\right| < \frac{1}{3} \] and recalling
that $|a_i|\leq 1$ we have \beqa \left| \frac{(
\sqrt{1-\epsilon_i}\sqrt{\epsilon_k}\, a_1 +
\sqrt{1-\epsilon_k}\sqrt{\epsilon_i}\, a_2 +
\sqrt{\epsilon_i}\sqrt{\epsilon_k} \, a_3  )}{\mu}\right| \nonumber
\\ \leq \frac{( \sqrt{1-\epsilon_i}\sqrt{\epsilon_k} +
\sqrt{1-\epsilon_k}\sqrt{\epsilon_i} +
\sqrt{\epsilon_i}\sqrt{\epsilon_k}   )}{\mu} \leq 1.\nonumber \eeqa
Also $\sqrt{\epsilon_i}\sqrt{\epsilon_k} \leq
\frac{\sqrt{\epsilon_i}+\sqrt{\epsilon_k}}{2}$ and $\mu \geq \eta$
for all non-orthogonal $\ket{\sigma_i}$ and $\ket{\sigma_k}$. Putting
all this together with eq. (\ref{mess}) we see that for all
non-orthogonal pairs $\ket{\sigma_i}$ and $\ket{\sigma_k}$ we have
\beq\label{zeta} |\inner{\beta_i}{\beta_k}| \geq 1-
\frac{3(\sqrt{\epsilon_i}+\sqrt{\epsilon_k})}{2\eta} \equiv 1-\zeta '
\eeq and note that $\zeta ' = O(\sqrt{\epsilon}) $. Hence we can
write \beq \label{bibk} \ket{\beta_k} = (1-\zeta) \ket{\beta_i} +
\sqrt{2\zeta - \zeta^2} \ket{\beta_{ik}^\perp} \eeq where
$\inner{\beta_i}{\beta_{ik}^\perp} =0$ and $\zeta =
O(\sqrt{\epsilon})$ so $\ket{\beta_k} \rightarrow \ket{\beta_i}$ as
$\epsilon \rightarrow 0$.

Recalling eq. (\ref{XI}) and using eq. (\ref{bibk}) in eq. (\ref{XK})
we have \beqa \nonumber \ket{\xi_i} & = & \sqrt{1-\epsilon_i}
\ket{\sigma_i}\ket{\beta_i} + \sqrt{\epsilon_i} \ket{\gamma_i}\\ &
\equiv \label{yuk1}& \sqrt{1-\epsilon_i} \ket{\sigma_i}\ket{\beta_i}
+ O(\sqrt{\epsilon}) \\ \nonumber \ket{\xi_k} & = &
\sqrt{1-\epsilon_k} (1-\zeta) \ket{\sigma_k}\ket{\beta_i} +
\sqrt{\zeta}\sqrt{(2-\zeta)} \sqrt{1-\epsilon_k}
\ket{\sigma_k}\ket{\beta_{ik}^\perp} +\sqrt{\epsilon_k}
\ket{\gamma_k}\\  & \equiv  & \sqrt{1-\epsilon_k}
\ket{\sigma_k}\ket{\beta_i} +O(\sqrt{\epsilon}) \label{yuk2} \eeqa
Now write
\[ \tr_A \proj{\xi_i} = \Omega_i . \]
Let us modify $\ket{\xi_i}$ into $\ket{\xi_i'}$ by replacing the
basis $\{ \ket{\sigma_i}, \ket{\tau_1}, \ket{\tau_2}, \ldots \}$ in
eqs. (\ref{XI}) and (\ref{gtau}) by the corresponding basis $\{
\ket{\sigma_k}, \ket{\tau_1'}, \ket{\tau_2'}, \ldots \}$ used in the
expression for $\ket{\xi_k}$. Then $\ket{\xi_k}$ is a purification of
$\Omega_k$ and $\ket{\xi_i'}$ is still a purification of $\Omega_i$.
Thus for all $\inner{\sigma_i}{\sigma_k} \neq 0$, we have
\[ F(\Omega_i , \Omega_k ) \geq |\inner{\xi_i'}{\xi_k}|^2 \geq
1-O(\sqrt{\epsilon} ). \] Now $\ce$ with $K$ states is irreducible so
by lemma 7 there is a chain of length at most $K$ from
$\ket{\sigma_{i_0}}$ to every other $\ket{\sigma_j}$. Hence for every
$j$, $\Omega_j$ is near to any chosen $\Omega_{i_0}$ in the following
sense: \beq\label{fido} F(\Omega_{i_0},\Omega_j ) \geq 1-K\,
O(\sqrt{\epsilon}). \eeq  We now compare the constant ensemble $
\ce_{const} = \{ \Omega_{i_0} ; p_i \}$ having $ \chi ( \ce_{const})
= 0 $, with the actual ensemble $ \ce = \{ \Omega_i ; p_i \}$ of
reduced states in register B arising from $\Gamma$, having $\chi
(\ce) = \chi $. From eq. (\ref{fido}) we have \[
\overline{F}(\ce_{const},\ce ) = 1-K O(\sqrt{\epsilon}) \] and the
result in appendix A gives $|\chi -0| \leq f(\epsilon )$ where
$f(\epsilon )$ has the form $A\sqrt{\epsilon}+B\sqrt{\epsilon}\log
\sqrt{\epsilon}$ for suitable constants $A$ and $B$, and  $f(\epsilon
) \rightarrow 0$ when $\epsilon \rightarrow 0$, as required.\, $\qed$

\noindent {\Large\bf Appendix C}

We use the notation $I$ to denote the index string $I= i_1 \ldots
i_n$ and $I_{\neq k}$ to denote the string $I$ with the $k^{\rm
th}$ position deleted. We aim to prove:

\begin{lemma} Let $p_I = p_{i_1}^{(1)} \ldots p_{i_n}^{(n)}$
be any product
distribution of $n$ probability distributions and let $\{ \rho_I ;
p_I \}$ be any associated ensemble of quantum states. Then
\begin{equation}
\chi( \{\rho_I;p_I \} ) \leq n \max_k \sum_{I_{\neq k}} p_{I_{\neq
k}}
    \left[ S( \sum_{i_k} p_{i_k} \rho_I )
        - \sum_{i_k} p_{i_k} S( \rho_I ) \right]
\end{equation}
\end{lemma}
We begin by defining the conditional and mutual entropies for a
quantum state $\sigma_{ABC}$ on three systems $A$, $B$ and $C$:
\begin{eqnarray}
S(A|B) &=& S(A,B) - S(B) \\ S(A:B) &=& S(A) + S(B) - S(A,B) \\
S(A:B|C) &=& S(A|C) + S(B|C) - S(A,B|C).
\end{eqnarray}
Here $S(A)$, $S(A,B)$ etc. denote the von Neumann entropies of the
states of the designated subsystems, obtained by partial trace
from $\sigma_{ABC}$. The following chain rules for conditional and
mutual  entropies are then simple consequences of the definitions:
\begin{eqnarray}
S(A_1,A_2,\ldots,A_n|B) &=& \sum_k S(A_k|A_{<k},B) \\
S(A_1,A_2,\ldots,A_n:B) &=& \sum_k S(A_k:B|A_{<k}).
\end{eqnarray}
(where $A_{<k}$ denotes the list $A_1 , \ldots ,A_{k-1}$). Now
suppose we are given a state $\sigma_{A_1 A_2 \dots A_n B}$ such
that $\sigma_{A_1 A_2 \dots A_n} = \sigma_{A_1} \otimes \dots
\otimes \sigma_{A_n}$.  We can calculate, for example, that
\begin{eqnarray}
S(A_2:B|A_1) &=& S(A_2|A_1) + S(B|A_1) - S(A_2,B|A_1) \\ &=&
S(A_1) + S(A_1,B) - S(A_1) - S(A_1,A_2,B) + S(A_1) \\ &=&
S(A_2:A_1,B).
\end{eqnarray}
 This
relationship can then be used to prove an upper bound on the joint
quantum mutual entropy as in the proposition below. This bound is
a quantum analogue of a classical mutual information inequality
given in \cite{biham}.
\begin{proposition}  For a state $\sigma_{A_1 A_2 \dots A_n B}$ such
that $\sigma_{A_1 A_2 \dots A_n} = \sigma_{A_1} \otimes \dots
\otimes \sigma_{A_n}$ the following inequality holds:
\begin{equation}
S(A_1,A_2,\ldots,A_n:B) \leq n \max_k S(A_k:A_{\neq k}, B).
\end{equation}
\end{proposition}

{\bf Proof}
\begin{eqnarray}
S(A_1,A_2,\ldots,A_n:B) &=& \sum_k S(A_k:B|A_{<k}) \\ &=& \sum_k
S(A_k:A_{<k}, B) \\ &=& \sum_k S(A_k: \tr_{>k} A_{\neq k}, B) \\
&\leq& \sum_k S(A_k: A_{\neq k}, B) \\ &\leq& n \max_k
S(A_k:A_{\neq k}, B).
\end{eqnarray}
The next to last inequality follows from the strong subadditivity
of von Neumann entropy, which implies that the quantum mutual
entropy cannot increase under any CPTP map. $\qed$

To obtain our desired inequality for $\chi$, we specialize to the
case where $A$ is a classical system correlated with $B$. Let
\begin{equation}
\sigma_{A_1 A_2 \dots A_n B} = \sum_I p_I \proj{I}_A \otimes
\rho_B^I
\end{equation}
where $I = i_1 i_2 \ldots i_n$, $p_I = \prod_{k=1}^n
p_{i_k}^{(k)}$ and $\ket{I} = \otimes_{k=1}^n \ket{i_k}_{A_k}$ for
sets of orthogonal states $\{ \ket{i_k}_{A_k} \}$.  Then a
straightforward calculation gives
\begin{eqnarray}
S(A_1,A_2,\ldots,A_n:B) &=& S(\sum_I p_I \rho_I) - \sum_I p_I
S(\rho_I) \\ &=& \chi( \{p_I;\rho_I \} )
\end{eqnarray}
and applying the proposition gives
\begin{equation}
\chi( \{\rho_I;p_I \} ) \leq n \max_k \sum_{I_{\neq k}} p_{I_{\neq
k}}
    \left[ S( \sum_{i_k} p_{i_k} \rho_I )
        - \sum_{i_k} p_{i_k} S( \rho_I ) \right],
\end{equation}
as required.

\end{document}